\documentclass[aps,prl,twocolumn,superscriptaddress]{revtex4}

\usepackage{graphicx}
\usepackage{dcolumn}
\usepackage{bm}
\usepackage{amsmath}

\begin{document}

\preprint{APS/123-QED}

\title{Formalism of optical coherency in material media with a quantum mechanical treatment}

\author{Ertan Kuntman}
 \affiliation{Departament de F\'isica Aplicada, Feman Group, IN2UB, Universitat de Barcelona, C/ Mart\'{\i} i Franqu\`es 1, Barcelona 08030, Spain.}
\author{M. Ali Kuntman}%
\affiliation{%
 Independent researcher, Ankara, Turkey\\
}%


\author{Jordi Sancho-Parramon}
\affiliation{
Rudjer Boskovic Institute, Bijenicka cesta 54, 10000, Zagreb, Croatia.\\
}%
\author{Oriol Arteaga}
\email{oarteaga@ub.edu}
\affiliation{Departament de F\'isica Aplicada, Feman Group, IN2UB, Universitat de Barcelona, C/ Mart\'{\i} i Franqu\`es 1, Barcelona 08030, Spain.}

\date{\today}

\begin{abstract}
The fluctuations or disordered motion of the electromagnetic fields are described by statistical properties rather than instantaneous values. This statistical description of the optical fields is underlying in the Stokes-Mueller formalism that applies to measurable intensities. However, the fundamental concept of optical coherence, that is assessed by the ability of waves to interfere, is not treatable by this formalism because it omits the global phase. In this work we show that, using an analogy between deterministic matrix states associated to optical media and quantum mechanical wavefunctions, it is possible to construct a general formalism that accounts for the additional terms resulting from the coherency effects that average out for incoherent treatments. This method generalizes further the concept of coherent superposition to describe how deterministic states of optical media can superpose to generate another deterministic media state. Our formalism of coherency is used to study the combined polarimetric response of interfering plasmonic nanoantennas.
\end{abstract}

\pacs{Valid PACS appear here}
\maketitle



In optics, interference is the phenomena that occurs when two coherent waves superpose. The celebrated example is the Young's double slit experiment with a beam of light, but quantum coherence and interference is
not restricted to photons. Any moving particle is susceptible to interfere with another if they keep a well-defined and
constant phase relation, as it can occur for example in between two oscillating dipoles \cite{book1}. In optics, this is one of the most fundamental interactions. When a material medium is irradiated by an electromagnetic wave, molecular electric charges are set in oscillatory motion by the electric field of the wave, producing secondary radiation in a form of refracted, reflected, diffracted or scattered light with certain polarization attributes. 


In quantum mechanics, the observable values are  the eigenvalues of Hermitian operators associated to the observable quantity. The observable corresponding to the optical phenomena occurring in light-matter interactions is the 4$\times$4 scattering matrix with sixteen real elements also known as the Mueller matrix that describes the linear transformation of the Stokes parameters of a light beam upon interaction with a linear medium. In this work, we first demonstrate how alternative representations of nondepolarizing (deterministic) optical systems that were recently presented \cite{KKA1} can be used to make the analogy between the scattering matrix \emph{states} of optical systems and the quantum mechanical wavefuction. We also show that quantum coherence in
material media can be represented by a coherent linear superposition of matrix (or vector) states associated to non-depolarizing Mueller matrices. This linear combination is generally understood as a convex sum of Jones matrices of nondepolarizing component systems \cite{Gil2007, Parke1949}. But here, instead of Jones matrices, we propose a linear combination of matrix (or vector) \emph{states} with \emph{complex} coefficients that play the role of probability amplitudes of quantum mechanics.  Despite the relationship between polarization optics and quantum mechanics has been studied in several previous works \cite{Fano1949, Fano1954, Wolf2013}, Mueller matrices have never been treated quantum mechanically.  Ossikovski \textit{et al.} \cite{Ossikovski2016} recently presented a treatment of spatial coherency in polarimetry and ellipsometry with Mueller matrices, albeit their formulation is based on classical electromagnetic first principles. In general, available theories about coherence and polarization \cite{gori, wolf2} require a direct consideration of electromagnetic fields. Nevertheless, our formalism is entirely based on phenomenological description of the polarized light using a quantum mechanical treatment. This generalized optical coherency formalism provides for the first time a direct and complete analogy between the Stokes-Mueller formalism describing interaction of light with the material medium and quantum mechanics.

The overall effect of the interaction of light with a deterministic, i.e., non-depolarizing, medium or optical system can be described by a 2$\times$2 complex matrix $\mathbf{J}$, referred to as Jones matrix \cite{Jones:41}. The 4$\times$4 real matrix for manipulating the Stokes vectors is the  Mueller matrix $\mathbf{M}$ that is directly connected with the experimental work in polarization optics. If the medium is deterministic then the associated Mueller matrix (also known as Mueller-Jones matrix) can be analytically obtained from the Jones matrix \cite{goldstein}. As opposed to the Jones matrix, a Mueller matrix does not contain information about the overall phase change introduced by a material medium, because it is not an observable.

Sometimes it is convenient to study the properties of a general Mueller matrix $\mathbf{M}$ (nondepolarizing or depolarizing) by transforming $\mathbf{M}$ into a Hermitian matrix $\mathbf{H}$ which is called the covariance matrix \cite{Cloude}. If and only if the Mueller matrix of the system is nondepolarizing, the associated covariance matrix will be of rank 1. In this case, it is always possible to define a covariance vector $|h\rangle$ such that
\begin{equation}
\mathbf{H}=| h\rangle\langle h|,
\end{equation}
where the vector $|h\rangle$ is the eigenvector of $\mathbf{H}$ corresponding to the single non-zero eigenvalue \cite{Gil2013, Gil2014, Aelio}. 

As its mathematical form suggests, $\mathbf{H}=| h\rangle\langle h|$ is an analog of the pure state of quantum mechanics expressed in the density matrix form, where the covariance vector, $|h\rangle$, plays the role of quantum mechanical state vector, $|\mathbf{\Psi}\rangle$. In a suitable basis that we have defined in a previous work \cite{KKA1}, the dimensionless components of $|h\rangle$ are  $\tau$, $\alpha$, $\beta$ and $\gamma$:
\begin{equation}\label{vectorh}
|h\rangle=\begin{pmatrix}
 \tau, \alpha, \beta, \gamma
\end{pmatrix}^T.
\end{equation}
where $\alpha$, $\beta$, $\gamma$ are complex parameters and $\tau$ can be chosen to be real because the global phase is not an experimental observable.

A deterministic state can be alternatively given by a Jones matrix $\mathbf{J}$, a Mueller-Jones matrix $\mathbf{M}_J$, covariance vector $|h\rangle$ or a $4\times4$ complex matrix, $\mathbf{Z}$, defined as \cite{KKA1}:
\begin{equation}
\mathbf{Z}=\begin{pmatrix}
\tau&\alpha&\beta&\gamma\\
\alpha&\tau&-i\gamma&i\beta\\
\beta&i\gamma&\tau&-i\alpha\\
\gamma&-i\beta&i\alpha&\tau
\end{pmatrix}.
\end{equation}
This matrix has a remarkable property \cite{KKA1}:
\begin{equation}
\mathbf{M}_J= \mathbf{Z}\mathbf{Z^*}=\mathbf{Z^*}\mathbf{Z},
\end{equation}
where $\mathbf{M}_J$ is a Mueller-Jones matrix.

Here, the analogy between the $\mathbf{Z}$ matrix and a quantum mechanical \emph{wavefuntion}, usually denoted as $\psi$, is evident. The $\mathbf{Z}$ matrix is a complex matrix state that, when multiplied with its complex conjugate, gives a real valued Mueller-Jones matrix with elements that are observable quantities in experimental polarization optics. In the following, $\mathbf{Z}$ matrices will be referred to as Mueller-Jones states, and we will show that it is also possible to think of a linear superposition of $\mathbf{Z}$ matrices in a way very similar to the superposition of quantum mechanical wavefunctions.

The coherent  superposition of polarization states can be introduced with  Young's double slit experiment.  The wavefunction of the combined beam can be written as a linear superposition of wavefunctions of light emerging from each slit:
\begin{equation}
\psi=a\psi_{a}+b\psi_{b}.
\end{equation}
The phenomenon of interference of light comes into play if $\psi_{j}$ are, in all respects,  identical to each other except  relative phases. For example, if  $\psi_{b}= e^{i\phi}\psi_{a}$ ($ 0\leq\phi<2\pi$), and if we let $a=b=\frac{1}{\sqrt[]{2}}$, then the probability distribution function at a given detection point displays a typical $\cos\phi$ dependence:
\begin{equation}\label{inter2}
\begin{split}
\psi\psi^*&=
\frac{1}{2}\psi_{a}\psi_{a}^*(1+ e^{i\phi})(1+ e^{-i\phi})\\&=\psi_{a}\psi_{a}^*(1+\cos\phi),
\end{split}
\end{equation}
If we consider an extended detector, the probability density at the detector  will vary accordingly with the cosine term as a function of position, because the optical path (and therefore the value of $\phi$) changes with the detection point.  On the other hand, if we set a vertical and a horizontal polarizer before each slit, there will be no sign of interference at all. Thus, it is worth to remark that the lack of interference does not necessarily indicate absence of coherence.

An optical superposition process may take place during a light-matter interaction experiment. When a light beam simultaneously illuminates different parts of the material medium, each part having different optical properties, the light emerging from different parts, in general with different polarizations, may coherently recombine into a single beam.  If the studied material medium is effectively composed of several non-depolarizing (deterministic) systems, each system with a well defined Jones matrix, then the Jones matrix of the combined system is simply given by a linear combination of the Jones matrices of the component systems\cite{Gil2007,Parke1949}:
\begin{equation}\label{unop}
\mathbf{J}=\sum_i\mathbf{J}_i,
\end{equation}
For the analogies with quantum mechanics that we are tracing in this work it is more practical to rewrite \eqref{unop} with normalized Jones component matrices, so that each term of the superposition is preceded by a complex coefficient that accounts for the relative amplitude and phase:
\begin{equation}\label{uno}
\mathbf{J}=a\mathbf{J}_{a}+b\mathbf{J}_{b}+ c\mathbf{J}_{c}+\cdots.
\end{equation}
By means of definition $\mathbf{Z}=\mathbf{A}(\mathbf{I}\otimes\mathbf{J})\mathbf{A}^{-1}$ (where $\mathbf{A}$ is a constant unitary matrix\cite{KKA1}), this coherent linear combination can be directly translated to the $\mathbf{Z}$ matrix states with the same complex coefficients:
\begin{equation}\label{dos}
\mathbf{Z}= a\mathbf{Z}_{a}+b\mathbf{Z}_{b}+ c\mathbf{Z}_{c}+\cdots.
\end{equation}
Complex coefficients  $a, b, c,...$, here play the role of probability amplitudes of quantum mechanics. Obviously, this is a coherent summation and the resultant matrix state $\mathbf{Z}$ corresponds to the nondepolarizing Mueller matrix  of the combined system. The complex coefficients, $a, b, c,...$, can generally be functions of space, time and frequency. These dependencies can entail depolarization effects if the measurement system cannot resolve these variations, as it will be discussed later.

Without loss of generality we may restrict our presentation to a two-term coherent parallel combination.  It can be shown that the Eqs. \eqref{uno} and \eqref{dos} lead to the same Mueller-Jones matrix of the combined nondepolarizing system, $\mathbf{M}_J$. For instance, from Eq. \eqref{dos}, $\mathbf{M}_J$ can be written in terms of $\mathbf{Z}$ matrices as follows:
\begin{equation}\label{tres}
\mathbf{M}_J=\mathbf{Z}\mathbf{Z}^*=aa^*\mathbf{Z}_a\mathbf{Z}_a^*+bb^*\mathbf{Z}_b\mathbf{Z}_b^*+ab^*\mathbf{Z}_a\mathbf{Z}_b^*+ba^*\mathbf{Z}_b\mathbf{Z}_a^*.
\end{equation}
In this expansion, $\mathbf{Z}_a\mathbf{Z}_a^*$ and $\mathbf{Z}_b\mathbf{Z}_b^*$ are the Mueller-Jones matrices of the nondepolarizing component systems, whereas, $\mathbf{Z}_a\mathbf{Z}_b^*$ and $\mathbf{Z}_b\mathbf{Z}_a^*$ are the matrices resulting from coherence that cannot be interpreted as Mueller matrices in the usual sense. The combined term $ab^*\mathbf{Z}_a\mathbf{Z}_b^*+ba^*\mathbf{Z}_b\mathbf{Z}_a^*$ turns out to be a real matrix; but, still it is not a Mueller matrix. The result provided by means of Eq. \eqref{uno} is mathematically equivalent to Eq. \eqref{tres} under the transformation $\mathbf{A}(\mathbf{J}_m\otimes\mathbf{J}_n^*)\mathbf{A}^{-1}=\mathbf{Z}_m\mathbf{Z}_n^*$ \cite{KKA1}. Besides rendering the mathematics compact and simple, the advantage of the $\mathbf{Z}$ matrix approach is that, in contrast to Jones formalism, it also  permits treating incoherent or partially coherent processes by, respectively, truncating or attenuating the coherence terms $\mathbf{Z}_a\mathbf{Z}_b^*$ and $\mathbf{Z}_b\mathbf{Z}_a^*$.

The Jones and the $\mathbf{Z}$ matrix approaches are equivalent descriptions for a coherent parallel combination of deterministic systems. However, sometimes it may be convenient to work with vectors rather than matrices,  and formulate the coherent parallel combination process in terms of the covariance vectors of the  associated  systems:
\begin{equation}\label{hequation}
|h\rangle=a|h_a\rangle+b|h_b\rangle+c|h_c\rangle+\cdots
\end{equation}
where, four dimensional complex vectors $|h_i\rangle$ are defined in Eq. \eqref{vectorh}.

In case of a two-term coherent parallel combination, the covariance matrix $\mathbf{H}$ of the combined system can be written as:
\begin{equation}
\begin{split}
\mathbf{H}=|h\rangle\langle h|=\\
aa^*|h_a\rangle\langle h_a|+bb^*|h_b\rangle\langle h_b|+ab^*|h_a\rangle\langle h_b|+ba^*|h_b\rangle\langle h_a|,
\end{split}
\end{equation}
where $|h_a\rangle\langle h_a|$  and $|h_b\rangle\langle h_b|$ are the covariance matrices corresponding to the Mueller-Jones matrices of the nondepolarizing component systems; $|h_b\rangle\langle h_a|$ and  $|h_b\rangle\langle h_a|$ are the mixed \emph{coherence} terms which cannot be related to the usual Mueller matrices. But, the covariance matrix of the combined system, $\mathbf{H}$, leads directly to the Mueller-Jones matrix of the combined system anyway.

In quantum mechanics, any state vector (pure state) can be written as a linear combination of basis states (pure states) which are, in general, a complete set of eigenvectors of a Hermitian operator that corresponds to an observable quantity:
\begin{equation}
|\mathbf{\Psi}\rangle=\sum_{i=1}^N a_i|\mathbf{\psi}\rangle_i
\end{equation}
where $a_i$ are complex numbers (amplitudes) and  $|\mathbf{\psi}\rangle_i$ are the eigenvectors of a Hermitian operator that constitute a complete set of basis system. The covariance vector $|h\rangle$ is analog of the quantum mechanical state vector, $|\mathbf{\Psi}\rangle$, and it also possible to decompose a given vector $|h\rangle$ with respect to a complete basis set of component systems. We simply apply the ordinary vector decomposition procedure:
\begin{equation} \label{superpositionh}
|h\rangle=a_1|h_1\rangle+a_2|h_2\rangle+a_3|h_3\rangle+a_4|h_4\rangle,
\end{equation}
where $a_i$ are complex coefficients  and $|h_i\rangle$ constitute a complete set of basis vectors. The vectorial decomposition of $|h\rangle$ is not unique: for a given $|h\rangle$ there may exist infinitely many decomposition with respect to different set of complete basis. Basis vectors, $|h_i\rangle$, can define an  orthogonal or  non-orthogonal basis. For example, if  $|h_1\rangle$ and $|h_2\rangle$ correspond, respectively, to orthonormal covariance vectors of a linear horizontal polarizer and a linear vertical polarizer, then the following expansion of $|h\rangle$ will correspond to a horizontal quarter-wave plate state:
\begin{equation} \label{combpol}
|h\rangle= \frac{1+i}{2}\,|h_1\rangle+\frac{1-i}{2}\,|h_2\rangle.
\end{equation}

Algebra of Mueller-Jones formalism admits a superposition of $|h\rangle$ states as given in Eq. \eqref{superpositionh}. Therefore, at least mathematically, we can consider an ideal quarter-wave plate state as a coherent linear combination of two orthogonal linear polarizer states. In practice, this means that, if it could be possible to combine two orthogonal polarizers coherently with the associated complex coefficients as given in Eq. \eqref{combpol}, we would obtain an artificial quarter wave plate that effectively responds to the incident light just like a genuine one. In general, we can use non-orthogonal basis to decompose a given covariance vector $|h\rangle$. However, decomposition with respect to non-orthogonal basis is more involved: we have to take into account covariant and contravariant types of vectors and expansion coefficients. As an example, the covariance vector of an ideal partial polarizer can be decomposed into a non-orthogonal basis states, one of them being the direct beam state which corresponds to the identity Mueller matrix, and the other component being a horizontal linear polarizer state, with suitable coefficients. 

In a real experiment, the measuring apparatus may be unable to resolve the fluctuations in the phases of the electromagnetic fields arising during the interaction of the light beam with a sample,  then the measured scattering matrix of the combined system turns out to be a depolarizing Mueller matrix that can be considered as a mixture of nondepolarizing Mueller-Jones matrices.  Kim, Mandel and Wolf \cite{Kim}, consider an \emph{ensemble} average of Jones matrix realizations in order to explain depolarization. Gil gives a more detailed depolarization scheme based on an \emph{incoherent} convex sum of Mueller-Jones matrices \cite{Gil2007}: if we let $I^{(i)}$ be the intensity of the portion of light that interacts with the ``$i$" element, and denote $\mathbf{J}^{(i)}$, $\mathbf{M}_J^{(i)}$ the respective Jones and Mueller-Jones matrices representing the ``$i$" element, the Jones vector ($\epsilon$) of the light pencil emerging from each element will be given by
\begin{equation}
\epsilon'_i= \mathbf{J}^{(i)}[\sqrt{p_i}\epsilon],
\end{equation}
where $p_i=I^{(i)}/I$, \:  $I$ being the total intensity. The corresponding Stokes vector,  $s'$,  of the complete emerging beam, obtained through the \emph{incoherent superposition} of the beams emerging from the different elements, $s'_i$, can be written as
\begin{equation}
s'=\sum_i s'_i=\left(\sum_i p_i \mathbf{M}_J^{(i)}\right)s=\mathbf{M}s,
\end{equation}
where $\mathbf{M}$ is the depolarizing Mueller matrix of the incoherently combined system.

In this result the system is considered as an ensemble, so that each realization ``$i$" characterized by a well-defined Mueller-Jones matrix $\mathbf{M}_J^{(i)}$, occurs with probability $p_i$, hence, the optical system can be considered as a proper mixture of Muller-Jones realizations at the outset. However, even when the fluctuations in phases in each one of the elements take place, instantaneous realizations are still deterministic. In other words, at a given time, space and frequency all phases can be considered as constants, therefore the linear superposition is \emph{instantaneously} coherent and the Mueller matrix of the combined optical system is instantaneously \emph{non-depolarizing} (here the adverb \emph{instantaneously} does not only imply a temporal meaning). Only when we begin to take into account the statistical averages (time average, spatial average and/or frequency average), coherence terms will be washed out and the result will be depolarizing. For example, consider a simple case where the $\mathbf{Z}$  matrix of the combined system is formed by a linear combination of $\mathbf{Z}$ matrices of two subsystems at a given instant:
\begin{equation}\label{similar}
\mathbf{Z}=\frac{1}{\sqrt{2}}\mathbf{Z}_a +\frac{e^{i\phi}}{\sqrt{2}}\mathbf{Z}_b,
\end{equation}
where $\mathbf{Z}_a$, $\mathbf{Z}_b$ are the matrix states of the subsystems, and $\phi$ is a constant phase angle. The nondepolarizing Mueller matrix corresponding to $\mathbf{Z}$ will be
\begin{equation}
\begin{split}
\mathbf{M}_J&=\frac{1}{2}\mathbf{M}_a +\frac{1}{2}\mathbf{M}_b+\frac{e^{-i\phi}}{2}\mathbf{Z}_a\mathbf{Z}_b^*+\frac{e^{i\phi}}{2}\mathbf{Z}_b\mathbf{Z}_a^*,
\end{split}
\end{equation}
where $\mathbf{Z}_a\mathbf{Z}_b^*$ and  $\mathbf{Z}_a^*\mathbf{Z}_b$ are the coherence terms. 

Now consider another instant:
\begin{equation}
\mathbf{Z'}=\frac{1}{\sqrt{2}}\mathbf{Z}_a -\frac{e^{i\phi}}{\sqrt{2}}\mathbf{Z}_b
\end{equation}
In this case there is an additional phase, $e^{i\pi}=-1$. Then the nondepolarizing Mueller matrix corresponding to $\mathbf{Z'}$ is
\begin{equation}
\begin{split}
\mathbf{M'}_J&=\frac{1}{2}\mathbf{M}_a +\frac{1}{2}\mathbf{M}_b-\frac{e^{-i\phi}}{2}\mathbf{Z}_a\mathbf{Z}_b^*-\frac{e^{i\phi}}{2}\mathbf{Z}_b\mathbf{Z}_a^*.
\end{split}
\end{equation}
In the arithmetic mean of $\mathbf{M}_J$ and $\mathbf{M'}_J$,
\begin{equation}
\mathbf{M}_{average} =\frac{1}{2}\mathbf{M}_a+\frac{1}{2}\mathbf{M}_b,
\end{equation}
 the coherence terms are totally truncated, and  the result is a \emph{depolarizing} Mueller matrix which turns out to be a convex sum of nondepolarizing Mueller matrices of the component systems.

The matrices $\mathbf{M}$ and $\mathbf{M'}$ are the instantaneous (in the sense of constant phase) realizations of the measurement process. Now consider a continuum of similar instantaneous realizations and assume that the phase relations between the component systems change very rapidly during the exposure time ($T$). For example, let the phase angle  $\phi$ be a function of time so that the orientation of unit vector $e^{i\phi}$  randomly fluctuates with a vanishing integral $\int_0^T e^{i\phi}\mathrm{dt}$, then, due to the temporal average of the instantaneous realizations, the coherence terms will be truncated (or attenuated in case of partial coherence) and depolarization effects will appear. Here we have discussed temporal averaging, but similar results would be obtained for spatial and frequency averaging.

This situation resembles to the development of an interference pattern on the screen of  Young's double slit experiment, photon by photon. The arrival of each photon at a point detector is an instantaneous realization of the superposed  probability waves. But, if the  coherency of light cannot be preserved in a long period of time, the interference pattern will be washed out, in spite of the fact that, the instantaneous detection of a single photon still obeys the well defined superposition principle of quantum mechanics. We may observe interference effects if in Eq. \eqref{similar} $\mathbf{Z}_a=\mathbf{Z}_b$:
\begin{equation}
\mathbf{Z}=\frac{1}{\sqrt{2}}\mathbf{Z}_a +\frac{e^{i\phi}}{\sqrt{2}}\mathbf{Z}_a =
\frac{1}{\sqrt{2}}\mathbf{Z}_a( 1+e^{i\phi}).
\end{equation}
This is an analog of Young's double slit with two equivalent component systems with a relative phase between them. The corresponding Mueller-Jones matrix is,
\begin{equation} \label{youngform}
\mathbf{M}_J=\mathbf{Z}_a\mathbf{Z}_a^*( 1+\cos\phi)=\mathbf{M}_a( 1+\cos\phi),
\end{equation}
where $\mathbf{M}_a$ is the nondepolarizing Mueller matrix (Mueller-Jones matrix) of the equivalent component systems. Note that Eq. \eqref{youngform} is an analog of Eq. \eqref{inter2}, but here interference effects are directly defined within Mueller matrices associated to optical media.

\begin{figure*}
\centering
\includegraphics[width=17cm]{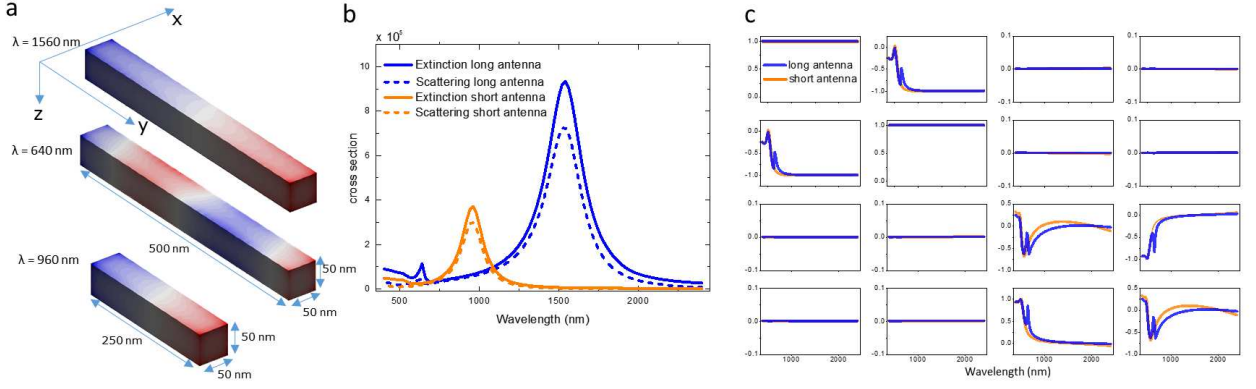}
\caption{\textbf{a}, Surface charge distributions (positive charges in red and negatives in blue) for the long and short nanoantennas at the indicated wavelengths. \textbf{b}, Calculated extinction and scattering cross sections for the two types of nanoantennas. Due to the different aspect ratio, the resonances occur at different wavelengths. \textbf{c}, Normalized Mueller matrix for each structure when they are vertically oriented (along y).}
\end{figure*}

Interference effects can only be observed if the value of $\phi$ can be preserved during the measurement. If $\phi$ varies drastically, on the average, $\cos\phi$ term will tend to vanish but the Mueller-Jones matrix of the combined system will be still equal to $\mathbf{M}_a$. For depolarization effects, uncontrollable-random fluctuations in the phases are not enough: at least two systems with \emph{distinct} states should be combined in parallel.

Superposition of distinct states can be illustrated by small (much smaller than the wavelength of light) spherical particles with isotropic polarizability that can be put in oscillatory motion when they are placed in a periodic electric field, producing secondary radiation. If, in an oriented material medium, dipoles are constrained to vibrate only along a certain direction, the forward scattering matrix of the dipole coincides with the Mueller matrix of a linear polarizer. Therefore, for vertical and horizontal dipoles we have, respectively:
   \begin{equation}\label{polarizer}
\mathbf{M_V} = \frac{1}{2}\left(
  \begin{array}{cccc}
   1 & -1  & 0 & 0 \\
   -1 & 1  & 0 & 0 \\
   0 & 0  & 0 & 0 \\
   0 & 0  & 0 & 0\\
  \end{array}
\right), \quad \mathbf{M_H} = \frac{1}{2}\left(
  \begin{array}{cccc}
   1 & 1  & 0 & 0 \\
   1 &  1 & 0 & 0 \\
   0 & 0  & 0 & 0 \\
   0 & 0  & 0 & 0\\
  \end{array}
\right).
\end{equation}
The corresponding $\mathbf{Z_V}$ and $\mathbf{Z_H}$ Matrices are
   \begin{equation}
\mathbf{Z_V} = \frac{1}{\sqrt{2}}\left(
  \begin{array}{cccc}
   1 & -1  & 0 & 0 \\
   -1 &  1 & 0 & 0 \\
   0 & 0  & 1 & i \\
   0 & 0  & -i & 1\\
  \end{array}
\right), \quad \mathbf{Z_H} = \frac{1}{\sqrt{2}}\left(
  \begin{array}{cccc}
   1 & 1  & 0 & 0 \\
   1 &  1 & 0 & 0 \\
   0 & 0  & 1 & -i \\
   0 & 0  & i & 1\\
  \end{array}
\right).
\end{equation}
The superposed state is given by $\mathbf{Z}=a_V\mathbf{Z}_V+a_H\mathbf{Z}_H$. If the two particles are identical and simultaneously excited by the same beam of light the complex weights $a_V$ and $a_H$ must be equal ($a_V=a_H=a$). Then, incoherent superposed state will be given by:
\begin{equation}
\mathbf{M}_{incoh}=|a|^2(\mathbf{Z}_V\mathbf{Z}^*_V+\mathbf{Z}_H\mathbf{Z}^*_H)=2|a|^2\left(
  \begin{array}{cccc}
   1 & 0  & 0 & 0 \\
   0 & 1  & 0 & 0 \\
   0 & 0  & 0 & 0 \\
   0 & 0  & 0 & 0\\
  \end{array}
\right),
\end{equation}
while for the coherent superposition:
\begin{equation}\label{identity}
\mathbf{M}_{coh}=|a|^2(\mathbf{Z}_V\mathbf{Z}^*_V+\mathbf{Z}_H\mathbf{Z}^*_H+\mathbf{Z}_V\mathbf{Z}^*_H+\mathbf{Z}_H\mathbf{Z}_V^*)=2|a|^2\mathbf{I},
\end{equation}
where $\mathbf{I}$ is the $4\times4$ identity matrix, meaning that the coherent superposed system is able to maintain the polarization state of any incoming beam. In fact, this is a general result when superposing $\mathbf{Z}$ matrices that correspond to orthogonal directions of anisotropy. For example the same identity matrix is recovered when superposing left- and right- handed circular polarizers.


If the particles are not identical or the applied periodic fields to each particle have different (but constant)
phases and amplitudes, the complex coefficients $a_V$ and $a_H$ may not coincide, and this will affect  $\mathbf{M}_{coh}$, term given by Eq. \eqref{identity}. On the other hand, regardless of the characteristics of the component particles, $\mathbf{M}_{incoh}$ will only be affected by the amplitudes of $a_V$ and $a_H$ but not by their phases.

The coherent superposition of dipoles can be well illustrated  for visible or near IR light by analyzing the optical response of thin stripes  of gold with the nanoantenna geometry shown in Fig. 1a.  These metallic rectangular structures have a width and thickness of 50 nm  and a length of 500 nm  (for long nanoantenna) and 250 nm (for short nanoantenna). The electromagnetic response of such antenna-like particles is calculated using the boundary element method (BEM) \cite{BEMGarcia, BEMHohenester}. We have used the MATLAB implementation of the BEM method developed by Hohenester \textit{et al.} \cite{BEMHohenester}.  The optical constants of of Au are taken from Johnson and Christy \cite{Johnson1972} with the data extrapolated to the infrared range by the Drude model. The extinction spectra of the long nanoantenna (Fig. 1b) shows a dipolar resonance at 1560 nm and a secondary quadrupolar resonance around 640 nm, as it is shown by the surface charge distributions of Fig. 1a. The short nanoantenna (Fig. 1b) has a single dipole resonance located at shorter wavelengths (960 nm), corresponding to a smaller aspect ratio \cite{BryantMap}. The simulated Mueller matrices for the vertically oriented short and long nanoantennas are shown in Fig. 1c. At long wavelengths, the Mueller matrix for both structures is very close to a vertical polarizer ($\mathbf{M}_V$ in Eq. \eqref{polarizer}), while at the shortest wavelengths energy of light is no longer confined in a dipolar resonance, and the nanoantennas behave more like a retarder.

\begin{figure}
\centering
\includegraphics[width=9cm]{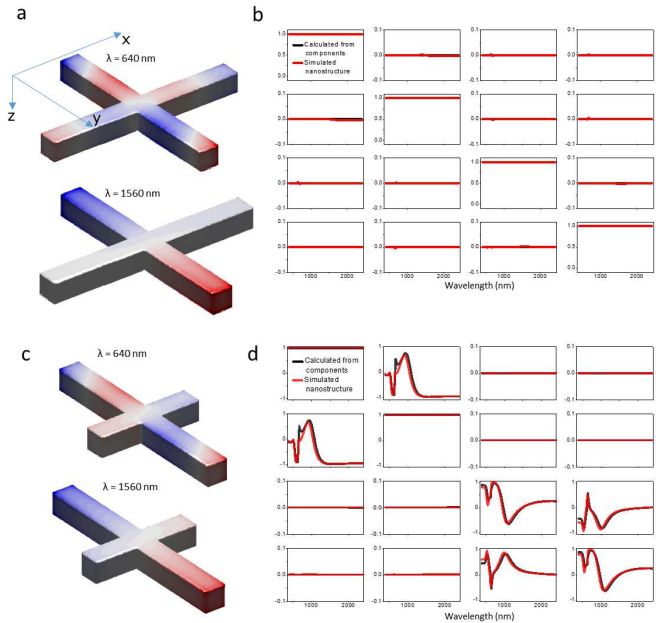}
\caption{ \textbf{a}, Surface charge distributions for a cross made of two  long nanoantennas at the indicated wavelengths. \textbf{b}, Comparison between the normalized Mueller matrix obtained from BEM simulation (red) of the structure in \textbf{a}, and the calculation  (black).   \textbf{c},  Surface charge distributions for a cross made of a long and a short nanoantenna. \textbf{d}, Comparison between the normalized Mueller matrix obtained from BEM simulation (red) of \textbf{c}, and the calculation (black).}
\end{figure}

In the next step, we analyze the superposed effect of two  combined nanoantennas that are not necessarily aligned. This combined effect can be calculated from Eq. \eqref{tres} by using the component $\mathbf{Z}$ matrices derived from the Mueller matrices of Fig. 1. We simply rotate the simulated Mueller matrices of vertical nanoantennas to obtain their Mueller matrix at an angle $\theta$: $\mathbf{R}(-\theta)\mathbf{M}\mathbf{R}(\theta)$. First we consider two perpendicularly crossed nanoantennas, which are illustrated by cross-like structures in Figs. 2a and 2c. For a cross formed by two equal nanoantennas, the complex coefficients associated to each component antenna are the same, then coherent superposition of orthogonal $\mathbf{Z}$ matrix states leads to an identity Mueller matrix (Fig. 2b), as it was anticipated by Eq. \eqref{identity}. However, even if the Mueller matrices of long and short nanoantennas are very similar (Fig. 1c), the combined effect  of perpendicularly crossed long and short antennas strongly differs from the identity mueller Matrix (Fig. 2d) because, in this case, the complex coefficients are not the same. For any of these perpendicularly crossed configurations, the Mueller matrices  simulated by the BEM method are in good agreement  with the matrices calculated from the data of component nanoantennas.

In a cross made by orthogonal nanoantennas there is no significant electronic interaction between the the dipole modes of the antennas and the extinction spectra is, qualitatively, an addition of the spectra of the individual antenna. However, the situation can be different if the dipole moments of the antennas are parallel or partially parallel because, in this case, they can significantly couple to each other. According to the plasmon hybridization theory of particle dimers, coupling of the individual resonances results into a lower energy mode with the dipole moments of the individual particles being in phase, and results into a higher energy mode with the dipole moments out of phase \cite{nordlander2004}. This second case has an overall lower dipole moment and hence scatters less light. The surface charge distribution calculated at the resonances  confirms the nature of these coupled modes (Fig. 1 of Supplementary Information).

\begin{figure*}
\centering
\includegraphics[width=17cm]{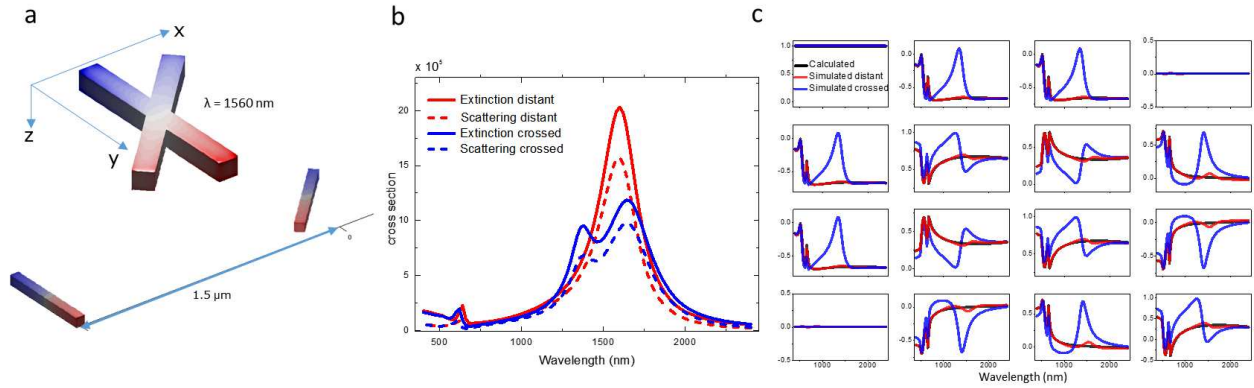}
\caption{ \textbf{a}, Surface charge distributions for crossed and distant (1.5 $\mu m$ of separation) oblique nanoantennas \textbf{b}, Extinction and scattering cross section for the systems shown in \textbf{a}. The strong coupling in the oblique crosses results into low and high energy modes that do not appear for the distant dipoles, when coupling is not significant. \textbf{c}, The calculated Mueller matrix from the component nanoantenna states is in agreement with BEM simulation for distant dipoles.}
\end{figure*}

In Fig. 3 we consider the superposed effect of the nanoantennas with a relative orientation of 45$^{\circ}$. As in this configuration the dipole moments are oblique, coupled modes can appear to substantially modify the individual responses of the antennas. The intensity of the coupling depends on the distance between the antennas (Fig. 2 of Supplementary Information). When the coupling is significant, the calculated Mueller matrices (with Eq. \eqref{tres}) from the associated component nanoantenna matrix states do not match the BEM simulations of the combined nanostructure. However, instead of simulating crossed antennas, when we consider separated antennas, as  shown in Fig. 3, the results of the BEM simulations show good agreement with the coherent superposition calculations of Eq. \eqref{tres}, because the coupling effects are minimized. Extending  our interference formalism with incorporating specific dynamical laws of interaction will be the subject of future works.

In this letter it is shown that the coherent (constant phase) parallel combination of deterministic systems can be written as a linear combination of $\mathbf{Z}$ matrices with complex coefficients. When the component Mueller-Jones states are the same  but have different relative phases interference effects are expected to be observed. It is also shown that depolarization can arise from temporal, spatial or frequency  averaging over fluctuating and distinct Mueller-Jones matrices. If the parallel combination process is incoherent at the outset, this averaging totally cancels out the coherence terms, and the Mueller matrix of the combined system reduces simply to the convex sum of Mueller-Jones matrix realizations.

The mathematical formalism we have described is based on the linear light-matter interactions described in Mueller matrices. It allows to introduce the concept of ``superposition of Mueller-Jones states" of optical media, and makes an analogy between the quantum mechanical wavefunction $\psi$ and the matrix state $\mathbf{Z}$. This constitutes a quantum theory of optical coherence that has the particularity that is grounded on the sixteen observable quantities (elements of a Mueller matrix) that characterize an optical media, as opposed  to the single observable quantity (intensity of light) around which other theories are built. Note that, despite the main subject being the optical coherence, this formalism does not directly entail working with electromagnetic fields.  We think that this formalism can be specially useful for applications in which coherent, partially coherent and incoherent superposition processes coexist. For example in nanophotonics it may provide theoretical means to tailor the light emission of nanostructures embedded in large area domains with the desired polarization response and functionality.

\bibliography{bibliography}

\end{document}